# Cognitive Depletion in the Wild: a Case Study of NMR Spectroscopy Analysis

**Lyndsey Franklin**
Pacific Northwest National Laboratory
Richland, WA, USA
lyndsey.franklin@pnnl.gov

**Nathan Hodas**
Pacific Northwest National Laboratory
Richland, WA, USA
nathan.hodas@pnnl.gov

## Abstract
NMR spectroscopy analysis is a detail-oriented analytic feat that typically requires specific domain expertise and hours of concentration. This work presents an ethnographic-style study of this analysis process in the context of evaluating the symptoms of cognitive depletion. The repeated, non-trivial decisions required by and the time-consuming nature of NMR spectroscopy analysis make it an ideal, real-world scenario to study the symptoms of cognitive depletion, its effect on workflow and performance, and potential strategies for mitigating its deleterious effects.

## ACM Classification Keywords
H.1.2 User/Machine Systems; H.4.1 Workflow Management; H.5.m Information Interfaces and Presentation (e.g. HCI) Miscellaneous; J.4 Social and Behavioral Sciences

## Introduction
In related work, we proposed a model of cognitive depletion symptoms [7]. Cognitive depletion is a class of fatigue involving the loss of analytic precision and other comprised abilities, which the subject may not perceive through self-assessment. Our purpose was to establish a set of objective cues that could be observed and used for the detection of a cognitively depleted state in a person who has been at work completing a single task for an extended period of time. We have developed our model of symptoms of cognitive depletion based on a literature review and are now in the process of verification and validation. Our first step is the observation of an analysis process that requires long periods of concentrated effort and involves detailed decisions using specific domain knowledge. That is, we will study an analysis task that may lead a person to be cognitively depleted affecting both his/her own well-being and performance. Our ultimate goal is to develop a mechanism for recommending mitigation strategies to a cognitively depleted person in the form of workflow suggestions or breaks.

Our first set of observations was conducted with the participation of nuclear magnetic resonance (NMR) spectroscopy analysts. NMR Spectroscopy is a detail-oriented analytic feat that requires long periods of involved decision-making and reasoning. It provides information about the structure and abundance of molecules in a sample material and is often used to investigate specific chemical processes, such as cellular metabolism. An NMR spectrum is visually represented in analysis software as series of peaks and valleys in intensity, varying over frequency space. These peaks and valleys represent the superposition and interaction of the constituent compounds as revealed by the magnetic perturbation of the NMR process. Spectra analysis is a non-trivial, manual process and requires extended training to be qualified to perform.

In this work, we present the findings of our case-study of an NMR analyst working to analyze a set of 80 related samples. We begin with interviews of two experts to understand the basic NMR Spectroscopy process and identify potential mechanisms of cognitive depletion. We then relate the observations from our study to our cognitive depletion symptoms model for verification and validation. We conclude with improvements to our model and future research directions.

## Symptoms of Cognitive Depletion

In other work, we discuss our model of symptoms that may be used to identify cognitive depletion [7]. For reference, we repeat the list of symptoms here. For a full discussion, see our related work.

*Vigilance and Reactionary Symptoms*
Vigilance and reactionary symptoms include phenomena such as habituation to status alerts, reaction time increases, distraction, and inattention. These symptoms are most likely to manifest in conditions require active monitoring of real-time events.

*Physical and Motor Symptoms*
Physical and motor symptoms include drowsiness, exhaustion, clumsiness, etc. as well as unintentional mistakes providing input, data, or commands to computerized interfaces.

*Personal Judgment Symptoms*
Personal judgment symptoms include occasions where a person is not accurately accounting for the state or their task and is confused, rushing tasks, or taking longer to make decisions than usual.

*Strategy Symptoms*
Strategy symptoms manifest as disruptive workflows, failure to form or adjust strategies, and the inclusion of unrelated multi-tasked actions.

## Interviews with Analysts

We interviewed two experts in NMR analysis before beginning our observations. These interviews allowed the research team to familiarize themselves with the NMR process and environment so that observations could focus on details of the process. Our two experts were interviewed individually at their regular places of work. Both experts were responsible for conducting and analyzing data from NMR and mass spectrometry experiments with a combined 9 years of direct experience at their current position. Our experts were given informed consent forms prior to the start of the interview. The research team took notes by hand during the interview.

*Analysis Process*
The experts were asked about their typical process for analyzing samples. Both described a typical dataset consisting of 70-80 spectra to be individually analyzed. In general, the process to analyze a single spectrum took approximately one hour if the analyst has a template to begin working from. Otherwise, to begin analysis of a single spectrum without a template could take several hours.

Both experts independently described a similar two-step process for completing their analysis using the Chenomix Profiler software [4]. The first "pass" over a spectrum is used to quickly make preliminary identifications, note oddities in the sample, and export the resulting analysis to Excel for statistical analysis. The preliminary identifications involve matching the "peaks and valleys" of the sample with expected and easily recognized compounds but not making precise concentration matches. One expert stated that she focused on identifying single-peak compounds first before reviewing any multi-peak clusters present in a spectrum. The second "pass" over a sample involves manipulating concentrations of the previously identified compounds to match the peaks and valleys of the spectrum. Some assistance is available through the software via a 'Fit Automatically' feature. One expert stated that she always verified automatic fits because they were not necessarily reliable.

Once analysis has been completed for a sample, an Excel report is created which includes the list of identified compounds and their concentrations. Included in this report is a color encoding of the analyst's certainty of the identification from "very certain" to "uncertain" to "not at all certain". This rating is based on the analyst's subjective judgment rather than statistics or some other certainty method.

*Symptoms of Cognitive Depletion*
We did not explicitly ask our experts about any of our previously identified symptoms from our model during the interviews. However, we did take the opportunity to note how our experts described their analysis process for admissions of fatigue and workflow conditions that would either alleviate or contribute to their cognitive depletion.

Both experts described their workflow as "broken up" and requiring several days to complete. Each had other tasks they were responsible for, occasionally interrupting their analysis and requiring them to go to a separate building. One of our experts reported that her analysis sessions were typically 2 hours long and that she did feel tired if she spent a whole day doing data analysis. We note that for several studies of cognitive and mental fatigue, participants are sufficiently fatigued in sessions lasting anywhere from 20 minutes to 2 hours depending on the task [1] [3] [8] [10] [11] [12] [13] [14]. The other expert stated that she avoided spending her entire (standard 8 hour) day doing analysis because she found it "draining". She also noted that due to multiple research projects in progress at once, she occasionally would have to stop analysis for one of them for several days to prioritize another. She often had to block time in her schedule to make sure her analysis tasks were completed.

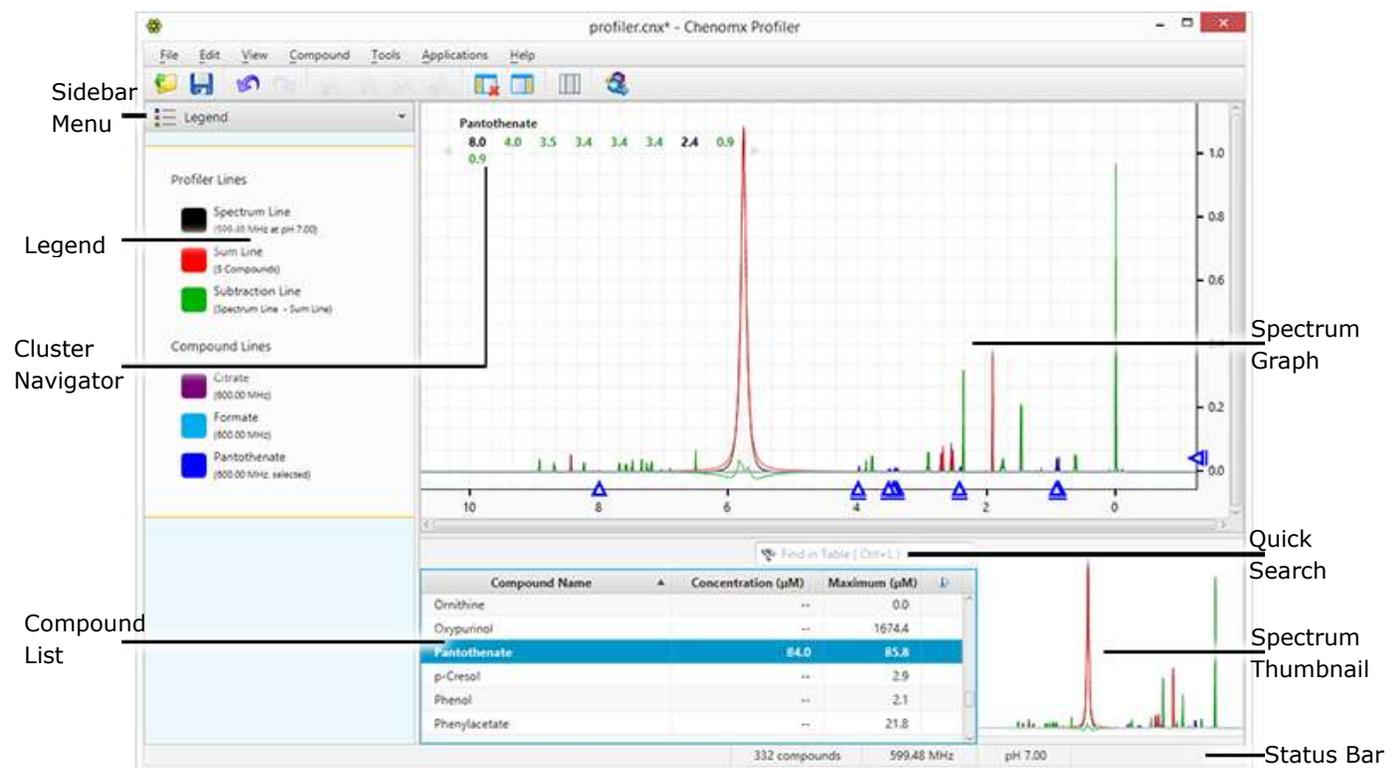

Figure 1. The Chenomix profiler used by our expert analyst during her work. From [4]

**Observing NMR Spectroscopy Analysis**
The next phases of our evaluation involved observing one of our experts perform NMR analysis as part of her normal duties. The samples she analyzed were all from a single data set assigned to the analyst as part of her regular work. As a result, our expert analyst was highly motivated to complete the analysis as accurately as possible.

> **Observed Symptoms of Cognitive Depletion**
>
> We observed or confirmed seven of our predicted symptoms of cognitive depletion. These included:
>
> - Task Rushing
> - Increased Decision Time
> - Command Errors
> - Distraction
> - Task Abandonment
> - Task Unrelated Thoughts
> - Physical Signs
>
> **Unobserved Symptoms**
>
> Certain predicted symptoms were not applicable to this case study based on the nature of the task and the expertise of the analyst. These included:
>
> - Confusion
> - Effort Over/Under Estimation
> - Habituation
> - Reaction Time Increases
> - Inattention
> - Information Inventory Control Failure
> - Strategy Inefficiency

*Work Environment*

Observations took place at our expert's regular place of work. This was a cubicle environment with approximately 8 other scientists occupying the same space. Our expert performed her work on a desktop computer with two monitors. The analyst used the Chenomix Profiler [4] software (figure 1), a standard software for such analysis (for example, see [9] [6] [2] [5]). A separate Excel spreadsheet was used to track which samples from the set of spectra to be analyzed had been completed, when, and by which analyst. Observing researchers took notes by hand on paper. Researchers brought with them a list of the symptoms from our model to track occurrences during the session. Additionally, we made use of the Tobii Pro Glasses 2 eye tracking system [15] to record the analysis session. Observers made no modifications to the analyst's regular work environment and gave the analyst no special instructions. She was asked to perform her analysis as she normally would were observers not present.

*Analysis in Action*

We observed the analyst as she completed work for 2.5 hours. In this time she refined the analysis of one previously analyzed sample and completed the preliminary analysis for an additional 5 samples from her assigned list of spectra. The majority of the analyst's time was spent working sequentially from an established list of compounds expected in the data set. For each compound, the analyst sequentially scanned each cluster peak of the compound and adjusted the position and height of peaks to approximately match the sum line of the sample spectrum. This involved manually clicking through the cluster navigation and manually adjusting the fit of compounds in the spectrum graph area. A context menu provided support options such as an 'automatic fit'.

*Symptoms of Cognitive Depletion*

Completing the analysis of the first sample took less than 10 minutes at the beginning of the session. Following this, the analyst began the preliminary analysis of new samples. The first two new samples of the analysis session were completed in 25 minutes and 23 minutes. The final three preliminary analyses took 20 minutes or less. In a post-session debriefing, the analyst admitted that she rushed the analysis of the last three samples we observed. In fact, the last sample was not completely analyzed and the analyst made notes in her Excel samples list that several compounds were still unidentified. This indicates both task rushing and task abandonment towards the end of the session, consistent with cognitive depletion.

Several distracting noises occurred due to the shared space of the laboratory when other scientists present left their workstations and returned for varying reasons. During the first 40 minutes of the session, the analyst showed no indication of distraction and analysis of eye-tracking data confirms that her attention remained on the profiler software. However, after 45 minutes the analyst began fidgeting in her chair whenever there was significant room noise. Eye-tracking data shows that during these periods the analyst fixated on whichever part of the screen she happened to be looking at for longer than usual compared with other moments when the room was quiet. This suggests the analyst could have been distracted by the noise, however the eye tracking data shows that the analyst remained focused on the profiler software even during moments of loud room noises.

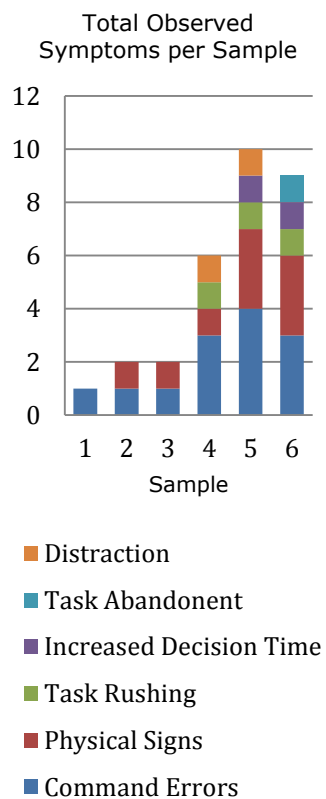

Figure 2. The total of observed symptoms per sample analyzed by the analyst during observations. This chart does not include symptoms discussed after the observation session.

The analyst received two email alerts during the observation session. Review of eye tracking data confirms that the second of these, occurring during analysis of the 3rd sample drew the analyst's attention (focus). However, the analyst otherwise ignored it and continued performing analysis.

Independent of room noise, the duration of fixation events increased in the last 40 minutes of the session. In a post-session debriefing, the analyst stated that none of the samples were more complex than the others and that each had between 40 and 50 compounds and that they were consistent with other samples from the data set she had previously analyzed. The increased fixation durations are thus suggestive of increased decision time related to cognitive depletion rather than complexity of the samples themselves.

Also independent of room noise, our analyst showed more signs of physical fatigue including fidgeting, sighing and yawning towards the end of the session. Observers noted that for the first 30 minutes of the session the analyst was in the same unmoving position (with the exception of her dominant, mouse-manipulating, right hand). The analyst yawned once during analysis of the second sample and fidgeted once during analysis of the third sample. The analyst yawned or fidgeted seven times total during analysis of the last three samples showing a notable increase in physical symptoms.

Finally, command errors were more frequent in the last 60 minutes of the session. The analyst did not use the undo command at all in the first three samples she analyzed. There were three uses of the undo command in total during analysis of the final two samples. Mis-clicking in the interface also became more common as the session progressed. Because the analyst worked sequentially through the list of compounds, it was obvious when compounds were clicked out-of-order and then corrected. There were three total mis-clicks in the compound list in analysis of the first three samples. In the last three samples, there were 3, 4, and 3 mis-clicks respectively.

**Discussion**

In our 2.5 hour observation, we observed six of our predicted symptoms from our model. During a post-session debriefing, our expert admitted that she did experience task-unrelated-thoughts occasionally during the session which brings our total reported or observed symptoms from the session to seven.

Several of our symptoms were not observed. The absence of habituation, reaction time increases, inattention, and information inventory control failure can all be explained by our expert analyst performing a task that did not require constant monitoring or real-time reaction to live events or streaming information. In fact, during our 2.5 hour observation session our expert analyst did not open or interact with any other programs besides the Chenomix software and the samples list in Excel (this may be due to being somewhat self-conscious in the presence of observers). The absence of confusion, effort over/under estimation, and strategy inefficiency can also be explained by the fact that our expert analyst was completing routine work on a common task for which she has already established an optimal workflow. Without external motivation, it is unlikely that she would experience a change in circumstances requiring her to adopt new workflow strategies beyond those taught to her during her training. Forgetfulness was not explicitly observed during the session nor was it mentioned in the post-session debriefing.

## Conclusion

Our initial interviews and observations indicate that our cognitive depletion symptoms model has captured several valid symptoms that can be detected in a real-world environment. Six of our predicted symptoms were directly observed and a seventh was confirmed during post-observation debriefings. Additional symptoms were not observed but are also not relevant to the task of NMR spectral analysis. We will be conducting more observations and interviews with additional experts to further validate our model. We will also conduct observations of analysis sessions lasting longer than the initial 2.5 hours discussed here to determine how many and when additional symptoms may manifest during the process. Ultimately, we will use these interviews and observations to inform improvements to our model and in turn use our model to develop a system which can detect these symptoms of cognitive depletion and recommend mitigation strategies before a person's performance and comfort decline.

## Acknowledgements

A portion of this research was performed onsite at EMSL, a DOE Office of Science User Facility sponsored by the Office of Biological and Environmental Research and located at Pacific Northwest National Laboratory.